\documentclass[letter]{aa} 
\usepackage{graphicx}
\usepackage{natbib}
\usepackage{txfonts}
\begin{document}

\titlerunning{Thermodynamics of Coronal Heating}

\title{Turbulent Coronal Heating Mechanisms: Coupling of\\ Dynamics and Thermodynamics}

\author{R. B. Dahlburg\inst{1} \and G.~Einaudi\inst{2} \and A. F. Rappazzo\inst{3} \and M. Velli\inst{4}}
	
\institute{	Laboratory for Computational Physics and Fluid Dynamics, 
         	      	Naval Research Laboratory, Washington, DC 20375, USA\\
		\email{rdahlbur@lcp.nrl.navy.mil}
	\and
		Berkeley Research Associates, Inc.,
		6537 Mid Cities Avenue, Beltsville, MD 20705, USA
	\and
		Bartol Research Institute, Department of Physics and Astronomy, University of Delaware, 
		DE 19716, USA
	\and
		Jet Propulsion Laboratory, California Institute of Technology, Pasadena, CA 91109, USA}

\date{\today} 

\abstract
{Photospheric motions shuffle the footpoints of the strong axial magnetic field that threads 
coronal loops giving rise to turbulent nonlinear dynamics characterized by the continuous formation 
and dissipation of field-aligned current sheets where energy is deposited at small-scales and the 
heating occurs.
Previous studies show that current sheets thickness is orders of magnitude smaller 
than current state of the art observational resolution ($\sim 700\, km$).} 
{In order to understand coronal 
heating and interpret correctly observations it is crucial to study the thermodynamics of such a system 
where energy is deposited at unresolved small-scales.}
{Fully compressible three-dimensional magnetohydrodynamic simulations are carried out to 
understand the thermodynamics of coronal heating in the magnetically confined solar corona.}
{We show that temperature is highly structured at 
scales below observational resolution and nonhomogeneously distributed so that only a fraction of the 
coronal mass and volume gets heated at each time.}
{This is a multi-thermal system where hotter and cooler 
plasma strands are found one next to the other also at sub-resolution scales and exhibit a temporal dynamics.}


\maketitle

\section{Introduction}

Magnetohydrodynamic (MHD) turbulence has long been implicated as a key process in
coronal heating \citep{ev96} as well as fast and slow solar wind 
acceleration.  Numerical simulations have been crucial in investigating the 
role of MHD turbulence.

To tackle the turbulence problem correctly with direct numerical simulations, it is 
necessary to resolve enough spatial scales to obtain an energy containing range, 
an inertial range, and a dissipation range. Because of storage needs, the more the 
complexity of the problem is reduced, the more spatial scales are obtainable.  
Hence most previous research in this area has been restricted to dynamical effects, 
e.g., reduced MHD (RMHD) \citep{s76} and cold plasma models.  This reduces the size of the 
problem considerably.  For example, for RMHD the dynamics of the system 
are determined by two variables: a stream function and a magnetic potential function 
\citep{rved07}.   For the cold plasma model it is  customary to evolve three velocity 
field components and three magnetic field components, neglecting the pressure term
(thus not correcting for the solenoidality of the velocity field) so that six variables are 
used \citep{dlkn09}.

In all these approximations all thermodynamical aspects are lost, i.e., there is no 
information on the connection between the dynamics of the system, responsible for 
the heating mechanism, and the temperature and density behavior. 
In particular energy lost through Ohmic and viscous diffusion is simply 
lost from the system and the thermodynamic physics involved with thermal conduction 
and optically thin radiation is neglected.

This lack of information prevents a complete reproduction of  the energy cycle:  
photospheric motions shuffle magnetic field-lines footpoints
injecting energy into the Corona and triggering non-linear dynamics
that form small-scales, organized in current sheets \citep{p72},where energy is then transformed
into thermal (and kinetic and perturbed magnetic) energy by means of magnetic reconnection.
Heat is then conducted from the high temperature corona back toward 
the low temperature photosphere where it is lost {\it via} optically thin radiation, 
but at the same time causing chromospheric evaporation.

Thermodynamics are typically studied with 1D hydrodynamic models
\citep{re10} that study either the equilibrium condition of a heated loop, or the time-dependent
evaporative and possible condensation flows. However they introduce an ad hoc heating function 
to mimic the energy deposition and neglect the cross-field dynamics (including field-lines reconnection).
It is therefore crucial to use the computationally more demanding three-dimensional compressible MHD 
(with eight variables), including an energy equation with thermal conduction and 
the energy sink provided by optically thin radiation.

RMHD turbulence simulations \citep{rved07,rved08} have shown that in the planes
orthogonal to the DC magnetic field the length of current sheets  is of the order of the
photospheric convective cells scale (i.e., $\sim 1,000\, km$), while their width is much
smaller and limited only by numerical resolution. In order to understand the thermodynamics
of such system in this letter we carry out simulations that resolve the scales below 
the convection scale.
The boundary conditions used in this paper do not allow evaporative flows to develop
along the heated loops, our focus here is on the relationship of turbulent current sheet 
dissipation and the forming coronal temperature structure.

Our new compressible code HYPERION has allowed us to make a start at examining the 
fully compressible, three-dimensional Parker coronal heating model.  HYPERION is a 
parallelized Fourier collocation Ð finite difference code with third-order Runge-Kutta time 
discretization that solves the compressible MHD equations with vertical thermal conduction 
and radiation included.  The results provided by this new compressible code preserve 
the spatial intermittency of the kinetic and magnetic energies seen in earlier RMHD simulations, 
but also provide new information about the 
evolution of the internal energy, i.e., we can now determine how the coronal plasma heats 
up and radiates energy as a consequence of photospheric convection of magnetic 
footpoints \citep{drv10}. 

In Section~2 we describe the governing equations, the boundary and initial conditions, 
and the numerical method.  In Section~3 we detail our numerical results.  Section~4 
contains a discussion of these results and some ideas about future 
directions of this research.

\section{Formulation of the problem}

\subsection{Governing equations}

We model the solar corona as a compressible, dissipative magnetofluid. The equations which govern such a system, 
written here in dimensionless
form, are:
\begin{eqnarray}
{{\partial n}\over {\partial t}}\ &=&\ -\nabla\cdot (n {\bf v}),\\
{{\partial n {\bf v}}\over{\partial t}}\ &=&\ -\nabla\cdot({n \bf v v}) 
   -{\beta}\nabla p + {\bf J}\times{\bf B}+
{1\over S_v}\nabla\cdot{\bf \zeta},\\ 
{{\partial T}\over{\partial t}}\ &=&\ -{\bf v}\cdot\nabla T  
 - (\gamma - 1) (\nabla\cdot {\bf v}) T
+\frac{1}{n}\big \{\frac{1}{ Pr\, S_v }\frac{\partial }{\partial z}(T^{\frac{5}{2}}\frac{\partial T}{\partial z}) 
\nonumber \\
&&\ +{(\gamma -1)\over\beta} [
{ 1\over S_v} \zeta_{ij} e_{ij}
+{1\over S} (\nabla\times{\bf B})^2 
-{1\over P_{rad} S_v} n^2\Lambda (T)
\nonumber \\
&&\ + {\beta\over(\gamma - 1)} n C_N]\big \},\\
{{\partial {\bf B}}\over{\partial t}}\ &=&\ \nabla\times{\bf v}\times{\bf B} 
  + \frac{1}{S}\nabla\times \nabla\times {\bf B}, \label{eq:b}
\end{eqnarray}
with the solenoidality condition $\nabla\cdot{\bf B} = 0$.  
The system is closed by the equation of state 
\begin{equation}
p = n T.
\end{equation}

The non-dimensional variables are defined in the following way:
$n ({\bf x}, t)$ is the numerical density,
${\bf v}({\bf x}, t) = (u, v, w)$ is the flow velocity,
$p({\bf x}, t)$ is the thermal pressure,
${\bf B}({\bf x}, t) = (B_x, B_y, B_z) $ is the magnetic induction field,
${\bf J} = \nabla\times{\bf B}$ is the electric current density,
$T({\bf x}, t)$ is the plasma temperature,
$\zeta_{ij}= \mu (\partial_j v_i + \partial_i v_j) - 
\lambda \nabla\cdot {\bf v} \delta_{ij}$ is the viscous stress tensor,
$e_{ij}= (\partial_j v_i + \partial_i v_j)$ is the strain tensor,
and $\gamma$ is the adiabatic ratio. To render the equations dimensionless we have used $n_0$, the photospheric 
numerical density, 
$V_A$, the vertical Alfv{\'e}n speed at the photospheric boundaries, 
$L_0$, the orthogonal box length (= $L_x$ = $L_y$) and $T_0$, the photospheric temperature.
Therefore time ($t$) is measured in units of Alfv\'en cross time ($\tau_A=L_0 /V_A$).

The function $\Lambda(T)$ that describes the temperature dependence of the radiation
is taken following~\cite{h74}, normalized with its value at the base 
temperature $T_0 = 10000\, K$.
The term $C_N$ denotes a Newton thermal term which is enforced 
at low temperatures to take care of the delicate chromospheric and transition region 
energy balance \citep{dn01}.
We use $C_N = 0.1~[T_i (z) - T(z)] e^{-2(z+0.5~L_z)}$ at the lower wall and
$C_N = 0.1~[T_i (z) - T(z)] e^{-2(0.5~L_z-z)}$ at the upper wall, 
where $T_i (z) $ is the initial temperature profile.

The important dimensionless numbers are:
$S_v = \rho_0 V_A L_0 / \mu \equiv$ viscous Lundquist number, 
$S = \mu_0 V_A L_0 / \eta \equiv$ Lundquist number,
$\beta = \mu_0 p_0 / B_0^2 \equiv$ pressure ratio at the wall,
$Pr = C_v \mu / \kappa T_0^{5/2} \equiv$ Prandtl number, and
$P_{rad} $, the radiative Prandtl number 
${\mu/ \tau_A^{2} n_0^2 \Lambda (T_0)} $, determines the strength of the radiation. $C_v$ is
the specific heat  at constant volume. 
The term "$\kappa$" denotes the thermal conductivity.
The magnetic resistivity $(\eta)$ and shear viscosity $(\mu)$
are assumed to be constant and uniform, and Stokes relationship is assumed  
so the bulk viscosity $\lambda = (2/3) \mu$. 

\subsection{Initial and Boundary Conditions}

We solve the governing equations in a Cartesian box of dimensions $1 \times 1 \times L_z$, 
where $L_z$ is the loop aspect ratio ($0\le x,y \le1,\ -L_z/2 \le z \le L_z/2$),
threaded by a DC magnetic field $B_0$ in the $z$-direction.
The system has periodic boundary conditions in $x$ and $y$,  and line-tied boundary 
conditions at the top and bottom $z$-plates where the vertical flows vanish ($v_z =0$)
and the forcing velocity in the $x$-$y$ plane is imposed as in previous studies 
\citep{hvhms96,ev96,ev99} evolving the stream function
\begin{equation}
\psi (x, y, t) = f_1 \sin^2 \left(\frac{\pi t}{2 t^*}\right) + f_2  \sin^2 \left(\frac{\pi t}{2 t^*} + \frac{\pi}{2}\right),
\end{equation}
where $f_i (x,y) =  \sum_{n,m} a_{nm}^i sin(k_n x + k_m y + \zeta_{nm}^i)$,
and  $\mathbf{v} = \nabla  \psi \times \mathbf{\hat e}_z$.
All wave-numbers with $3\le(k_n^2 + k_m^2)^{1/2}\le4$ are excited, so that 
the typical length-scale of the eddies is $\sim 1/4$. As the typical convective cell
size is $\sim 1,000\, km$ and we use $L_z = 5$ this implies that our computational 
box in conventional units spans $4,000^2 \times 20,000\, km^3$.
Every $t^*$, the coefficients $a_{nm}^i$ and the $ \zeta_{nm}^i$ are randomly 
changed alternatively for eddies 1 and 2. The magnetic field is expressed as  
$\mathbf{B} = B_0 \mathbf{\hat e}_z + \mathbf{b}$, where $\mathbf{A}$ is the vector 
potential associated with the fluctuating magnetic field 
$\mathbf{b} =  \nabla \times \mathbf{A}$. At the top and bottom $z$-plates $B_z$, n and T 
are kept constant at their initial values $B_0$, $n_0$ and $T_0$, while the magnetic vector potential 
is convected by the resulting flows.

In what follows we have assumed the normalizing quantities to be: $n_0=10^{17}\, m^{-3}$, 
$T_0=10^{4}\, K$, $B_0=10^{-2}$ tesla, and $L_0 = 4 \times 10^{6}\, m$. 
It follows that the values of the various parameters appearing in the equations  are: 
$\beta = 1.7 \times 10^{-4}$, $v_A = 6.9 \times 10^{5}\, m/s$, $\tau_A = 5.8\, s$, 
$S = 2.7\times10^{9}$, $S_v = 2.09 \times 10^{9}$, $Pr = 1.52 \times 10^{-2}$, 
$P_{rad} = 3.35 \times 10^{-6}$, $\kappa \times T_0^{5/2}=0.18\, W m^{-1} K^{-1}$,  
$\ln \Lambda = 10$.

The normalized time scale of the forcing, $t^*$, is set to 51.7 to represent 
the 5 minutes typical photospheric convection time scale. The normalized  
photospheric velocity is $V_0 = 1.45 \times 10^{-3}$, corresponding to $10^3\, m/s$.

We impose as initial conditions a temperature profile with a base temperature 
$T_0=10^4\, K$ and a top temperature $8\times10^5\, K$ with the following z-dependence
for the nondimensional temperature $T_i (z)  = 1 + 79 \cos( \pi z / L_z)$, while the 
numerical density is given as $n_i (z) = 1/T_i (z)$. 
This choice allows us to neglect the gravitational effects since for most of the 
loop the temperature remains high enough to make the
gravitational length-scale bigger than $L_z$ at all times.

\begin{figure}
\begin{centering}
\includegraphics[scale=.55]{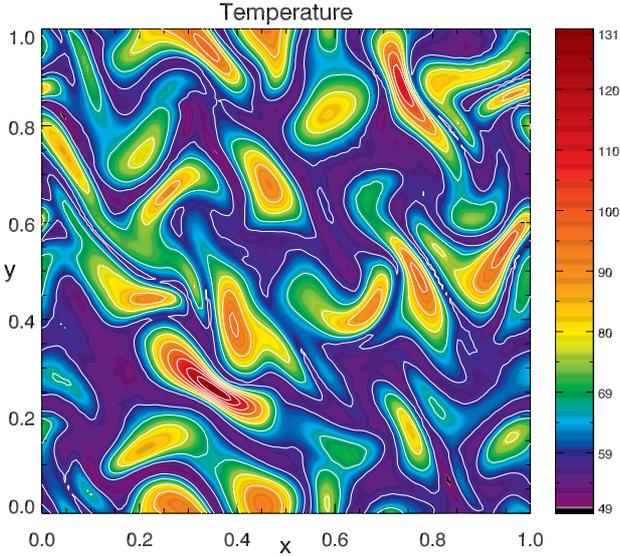}
\caption{Temperature contours at $t=300$ in the midplane ($z=0)$.
Maximum (130) and minimum (49) temperatures correspond to $1.3$ 
million and $490$ thousand degrees Kelvin in conventional units.
Distances are normalized by $L_0 = 4\times 10^6~m$.
\label{fig1}}
\end{centering}
\end{figure}

\section{Results}

With previous definitions equation~(\ref{eq:b}) can be replaced by the 
magnetic vector potential equation:
\begin{equation}
 {\partial {\bf A}\over\partial t}={\bf v}\times ( B_0\, \mathbf{\hat e}_z + \nabla\times{\bf A} )
  + {1\over S}~ \nabla\times\nabla\times {\bf A} 
\end{equation}
We solve numerically the equations~(1)-(3) and (7) together with equation~(5).  Space is discretized in $x$ and 
$y$ with a Fourier collocation scheme  \citep{dp89} with isotropic truncation dealiasing.  
Spatial derivatives are calculated in Fourier space, and nonlinear product terms are 
advanced in configuration space.  
A second-order central difference technique \citep{dmz86} is used for the 
discretization in $z$.  A time-step splitting scheme is employed.  
The code has been parallelized using MPI.  A domain decomposition is employed in which the computational 
box is sliced up into $x$--$y$ planes along the $z$ direction. 

We present the results obtained running the code with $S = S_v = 4\times10^4$
with a resolution of $128^3$.
In order to keep the efficiency of the radiative and conductive terms in the energy equation 
as in the real corona, we have rescaled $Pr$ and $P_{rad}$ accordingly with the choice 
of $S_v$, i.e., $Pr = 793.39$ and $P_{rad} = 0.175$. 
This choice is due to the result found in the RMHD model \citep{rved08}
that turbulent dissipative processes  are independent of viscosity and
resistivity when an inertial range is well resolved.

The system starts out in a ground state, threaded by a guide magnetic field:  the interior of 
the channel has zero perturbed velocity and magnetic fields. The footpoints of the magnetic 
field are subjected to convection at the $z$ boundaries with initial number density  and 
temperature profiles assigned as described above.  
The behavior of the volume averaged quantities, such as kinetic and fluctuating magnetic 
energies and resistive and viscous dissipation show a temporal behavior similar to the 
previous RMHD results obtained by \cite{rved07,rved08,rve10}.  

The fluctuating magnetic and kinetic energies 
($e_v ={1\over 2} \langle n {\mathbf v}^2\rangle$
and
$e_b={1\over 2} \langle {\mathbf b}^2\rangle$),
as shown in previous RMHD model, and
the total internal energy ($U=\langle n T\rangle {\beta / (\gamma - 1)}$), 
not obtainable in the RMHD case exhibit an intermittent behavior.
After the system is driven by photospheric motions for more than
$100$ Alfv\'en times a statistically stationary state is attained and all quantities
fluctuate around their steady values.  

\begin{figure}
\begin{centering}
\includegraphics[scale=.49]{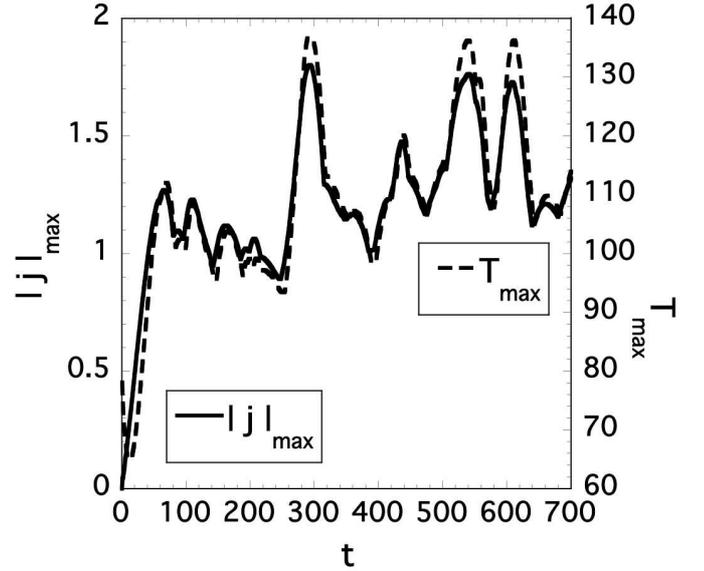}
\caption{Maximum temperature ($T_{max}$) and maximum electric current $|{\bf j}|_{max}$ as
functions of time.
\label{fig2}}
\end{centering}
\end{figure}

We have verified that the dynamics of this problem are
well represented by RMHD.
We define the volume-averaged velocity  $\theta_v$ and magnetic $\theta_b$ 
anisotropy angles as in \cite{smm83}.
For isotropic turbulence, $\theta_v$ and $\theta_b$ would equal $45^{\circ}$. 
For fully anisotropic turbulence, $\theta_v$ and $\theta_b$ would equal $90^{\circ}$, 
implying that their spectra would be normal to $B_0\, \mathbf{\hat e}_z$.
We find that, after $t=100$, $\theta_b$ is always bigger than $87^{\circ}$,
while the velocity field fluctuations are 
found to be slightly more anisotropic than those in the magnetic field, 
$88^{\circ} < \theta_v < 89^{\circ}$, 
indicating that the turbulence in the present system is highly anisotropic, i.e., the turbulence
is strongly confined to planes orthogonal to the guide magnetic field.
Moreover we have verified that the dynamics is almost 
incompressible.

The important new result of this letter refers to the thermodynamical behavior of the system
resulting from the self-consistent heating due to the magnetic field-line tangling induced 
by photospheric motions, once the cooling effects of conduction and radiation are taken into account. The fact that the total 
internal energy, as mentioned above, fluctuates around a steady value means 
that the self-consistent heating 
mechanism due to photospheric motions is efficient enough to sustain a hot corona once the system has reacted to 
photospheric motion. Such a reaction leads to the formation of small scales where energy can be efficiently dissipated. The 
small scales, corresponding to the presence of current sheets, are not uniformly distributed in the system, and as a result 
the temperature is spatially intermittent, as shown in Figure~\ref{fig1}, where we show the midplane temperature contours at 
$t=300$. 

Figure~\ref{fig2} shows the time evolution of   
$T_{max}$ and $|{\bf j}|_{max}$, the maximum temperature and the maximum current density present in the loop at each 
time respectively. It can be seen 
that the $T_{max}$ temperature peaks oscillate in time and, after an initial transient, the $T_{max}$ values are well above  
$T=10^6\, K$, higher than the initial peak temperature of  $T=8\times10^5\, K$ degrees.   
An important fact to emphasize is that the peaks in 
$T_{max}$ roughly coincide with bursts in the maximum electric current.
Since currents are organized in sheets elongated along the mean magnetic field
and the heating is due to the dissipation of currents we expect the heating 
not to be homogeneously distributed throughout the loop.
\begin{figure}
\begin{centering}
\includegraphics[scale=.345]{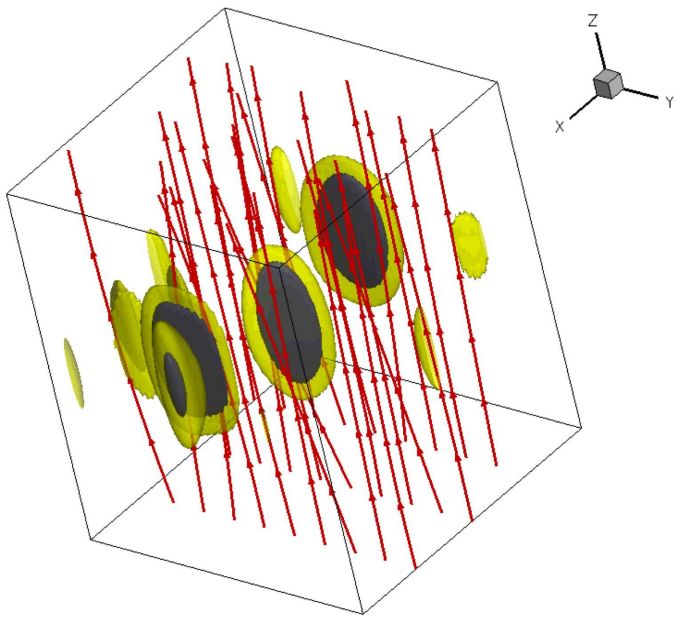}
\caption{Temperature is organized in quasi-2D pancake-like structures
aligned with the magnetic field.
The two isosurfaces for $T=10^6\, K$ and $T=1.2\times10^6\, K$ 
at time $t=300$ are shown respectively in yellow and grey
(corresponding to $T=100, 120$ in nondimensional units). 
Sample magnetic field-lines are drawn in red. The box has 
been rescaled to improve visualization.
\label{fig3}}
\end{centering}
\end{figure}

This expectation is confirmed by Figure~\ref{fig3}, 
where a 3-D snapshot of the isosurfaces of two temperatures ($T=10^6$ and 
$T=1.2\times10^6\, K$) at $t=300$ is presented. 
The spatial temperature morphology shows that the high temperatures are reached
in selected quasi-2D regions elongated along the mean magnetic field 
where the system dynamics places at each time the current sheets where energy is 
dissipated. The higher temperatures are found at the center of these structures 
(up to $1.35\times10^6\, K$ at $t=300$), while
farther out the temperature falls off to a background value that in the central z-region
averages around $6\times 10^5\, K$.
Snapshots taken at different times show similar behavior with different high temperatures 
locations.

\section{Conclusions and discussion}

In this paper we have presented for the first time the thermodynamic behavior of a 
system where the conductive and radiative losses are balanced by the heating process 
due solely to the photospheric motions and the subsequent tangling of magnetic field 
lines, resolving the scales below convection. While previous compressible simulation 
\citep{gn02,bp11} have considered entire active regions resolving at most the convective 
scale, it is pivotal to resolve smaller scales in order to understand how the plasma is
heated and its observational consequences.

A lot of work has been done in the past to show the reaction of a plasma  subject 
to mass flow at its boundaries. It has been shown that the system reacts developing 
a weak turbulent state, where energy is dissipated in the current sheets resulting 
from the dynamics induced by the boundary velocity forcing. The results of the present 
work show that this process leads to a loop where \emph{the temperature is spatially highly structured}.

A loop is therefore a multi-thermal system, where the temperature peaks around
current sheets and exhibits a distribution of temperatures. The characteristics of this
distribution (peak, width, background) will vary depending on the parameters of the 
loop (length, boundary velocity, mean magnetic field).

Temperature and heating are highly nonhomogeneously distributed so that
only a fraction of the loop volume exhibits a significant heating at each time.
Therefore emission observed in a spectral band origins from the small sub-volume
at the corresponding temperature, whose mass is considerably lower that the total
loop mass.

Observations currently have a spatial resolution of $\sim 700\, km$ and
a temporal resolution of $\sim 30\, s$. These temperature structures are 
therefore unresolved, and to interpret observations correctly the
spatial and temporal intermittency must be taken into account.
We intend to present in a following  paper how our loop appears once the emission 
is averaged over the coarser observational resolutions. At those scales the loop might result
multithermal or approximately isothermal depending on the loop parameters.

It is very likely that the local thermodynamical equilibrium is lost and that the 
emission is due to particles accelerated in the current sheets rather than direct 
transformation of magnetic energy into thermal energy. This is beyond the MHD model.

We have shown that the heating due to photospheric motions is able to balance the 
losses and that the dynamics induced by the such motions is a multi-scale dynamics 
where the energy is released in a number of spatial spots which account for the observed emission.


The density stratification  in our simulation allows limited 
evaporation from the z-boundary edges, where density is higher but does
not reach typical chromospheric values.
In future work we plan to implement characteristic boundary conditions
\citep[e.g.,][]{rved05}
that allow higher density flows through the $z$-boundaries,
that can also give rise to condensation and prominences, and study 
their impact on the thermodynamics.

\begin{acknowledgements}
This work was carried out in part at the Jet Propulsion Laboratory
under a contract with NASA. R.B.D.\ thanks 
Dr.~P.~Byrne for helpful conversations.
\end{acknowledgements}


\begin{thebibliography}{}
%
\bibitem[Bingert \& Peter(2011)]{bp11} Bingert, S., \& Peter, H.  2011,
A\&A, 530, A112
%
\bibitem[Dahlburg et al.(2009)]{dlkn09} Dahlburg, R. B., Liu, J.-H., Klimchuk, J. A., \& Nigro, G. 2009,
\apj, 704, 1059
%
\bibitem[Dahlburg, Montgomery \& Zang(1986)]{dmz86} Dahlburg, R. B., Montgomery, D.,
\& Zang, T. A. 1986,  J.\ Fluid Mech., 169, 71
%
\bibitem[Dahlburg \& Picone(1989)]{dp89} Dahlburg, R. B., \& Picone, J. M. 1989, 
Phys.\ Fluids B, 1, 2153
%
\bibitem[Dahlburg et al.(2010)]{drv10} Dahlburg, R. B.,  Rappazzo, A. F., \& Velli, M. 2010, 
AIP Conf. Proc., 1216, 40
%
\bibitem[Dorch \& Nordlund(1999)]{dn01} Dorch, S.B.F., \& Nordlund, \AA.  2001, A\&A, 365, 562
%
\bibitem[Einaudi \& Velli(1999)]{ev99} Einaudi, G., \& Velli, M.  1999, Phys.\ Plasmas, 6, 4146
%
\bibitem[Einaudi et al.(1996)]{ev96} Einaudi, G.,  Velli, M.,  Politano, H., \& Pouquet, A. 1996,
\apj,  457, L113
%
\bibitem[Gudiksen \& Nordlund(2002)]{gn02} Gudiksen, B. V., \& Nordlund, \AA. 2002, \apj,
572, L113
%
\bibitem[Hendrix et al.(1996)]{hvhms96} Hendrix, D. L., Van Hoven, G.,  Miki\'c, Z., \& Schnack, D. D. 
1996, \apj, 470, 1192
%
\bibitem[Hildner(1974)]{h74} Hildner, E. 1974, Sol.\ Phys., 35, 123
%
\bibitem[Parker(1972)]{p72} Parker, E. N.1972, \apj, 174, 499
%
\bibitem[Shebalin et al.(1983)]{smm83} Shebalin, J. V., Matthaeus, W. H., \& Montgomery,  D. 1983,
J.\ Plasma Phys., 29, 525
%
\bibitem[Rappazzo et al.(2005)]{rved05} Rappazzo, A. F.,   Velli, M.,  Einaudi, G., \& Dahlburg, R. B.
2005, \apj, 633,  474
%
\bibitem[Rappazzo et al.(2007)]{rved07} Rappazzo, A. F.,   Velli, M.,  Einaudi, G., \& Dahlburg, R. B.
2007, \apj, 657, L47
%
\bibitem[Rappazzo et al.(2008)]{rved08} Rappazzo, A. F.,   Velli, M.,  Einaudi, G., \& Dahlburg, R. B.
2008, \apj, 677, 1348
%
\bibitem[Rappazzo et al.(2010)]{rve10} Rappazzo, A. F.,   Velli, M.,  \& Einaudi, G. 2010,
\apj, 722, 65
%
\bibitem[Reale(2010)]{re10} Reale, F. 2010, Living Rev.\ Solar Phys.\, 7, 5
%
\bibitem[Strauss(1976)]{s76} Strauss, H. R.\ 1976, Phys.\ Fluids 19, 134
%
\end{thebibliography}
\end{document}